\documentclass[11pt,a4paper]{article}
 
\usepackage{amsmath,amsthm,amssymb}

\usepackage{cite}

\pdfoutput=1

\usepackage{graphics,graphicx}
\usepackage{epsfig}
\usepackage{multicol}
\usepackage{color}
\makeatletter
\@addtoreset{equation}{section}
\makeatother

\usepackage{braket}
\usepackage{bm}

\begin{document} 
\sloppy

\begin{center}
\LARGE{{\bf Generalised Uncertainty Relations from Finite-Accuracy Measurements}}
\end{center}

\begin{center}
\large{Matthew J. Lake${}^{a,b,c,d,e*}$\footnote{matthewjlake@narit.or.th}, Marek Miller${}^{f}$\footnote{m.miller@cent.uw.edu.pl}, Ray Ganardi${}^{f}$\footnote{r.ganardi@cent.uw.edu.pl} and Tomasz Paterek${}^{g,h}$}
\end{center}
\begin{center}
\emph{$^{a}$National Astronomical Research Institute of Thailand, \\ 260 Moo 4, T. Donkaew,  A. Maerim, Chiang Mai 50180, Thailand \\}
\emph{$^{b}$Department of Physics and Materials Science, \\ Faculty of Science, Chiang Mai University, \\ 239 Huaykaew Road, T. Suthep, A. Muang, Chiang Mai 50200, Thailand \\}
\emph{$^{c}$School of Physics, Sun Yat-Sen University, \\ Guangzhou 510275, People’s Republic of China \\}
\emph{$^{d}$Department of Physics, Babe\c s-Bolyai University, \\ Mihail Kog\u alniceanu Street 1, 400084 Cluj-Napoca, Romania \\}
\emph{$^{e}$Office of Research Administration, Chiang Mai University, \\ 239 Huaykaew Rd, T. Suthep, A. Muang, Chiang Mai 50200, Thailand \\}
\emph{$^{f}$Centre for Quantum Optical Technologies, Centre of New Technologies, \\ 
University of Warsaw, Banacha 2c, 02-097 Warsaw, Poland \\}
\emph{$^{g}$Department of Physics, Xiamen University Malaysia, 43900 Sepang, Malaysia \\}
\emph{$^{h}$Institute of Theoretical Physics and Astrophysics, \\ Faculty of Mathematics, Physics and Informatics, \\
University of Gda\'nsk, 80-308 Gda\'nsk, Poland}
\vspace{0.1cm}
\end{center}

\begin{abstract}
In this short note we show how the Generalised Uncertainty Principle (GUP) and the Extended Uncertainty Principle (EUP), two of the most common generalised uncertainty relations proposed in the quantum gravity literature, can be derived within the context of canonical quantum theory, without the need for modified commutation relations. 
A GUP-type relation naturally emerges when the standard position operator is replaced by an appropriate Positive Operator Valued Measure (POVM), representing a finite-accuracy measurement that localises the quantum wave packet to within a spatial region $\sigma_g > 0$. 
This length scale is the standard deviation of the envelope function, $g$, that defines the POVM elements. 
Similarly, an EUP-type relation emerges when the standard momentum operator is replaced by a POVM that localises the wave packet to within a region $\tilde{\sigma}_g > 0$ in momentum space. 
The usual GUP and EUP are recovered by setting $\sigma_g \simeq \sqrt{\hbar G/c^3}$, the Planck length, and $\tilde{\sigma}_g \simeq \hbar\sqrt{\Lambda/3}$, where $\Lambda$ is the cosmological constant. 
Crucially, the canonical Hamiltonian and commutation relations, and, hence, the canonical Schr{\" o}dinger and Heisenberg equations, remain unchanged. 
This demonstrates that GUP and EUP phenomenology can be obtained without modified commutators, which are known to lead to various pathologies, including violation of the equivalence principle, violation of Lorentz invariance in the relativistic limit, the reference frame-dependence of the `minimum' length, and the so-called soccer ball problem for multi-particle states.
\end{abstract}

\section*{}
{{\bf Keywords}: Generalised uncertainty relations, generalised uncertainty principle, extended uncertainty principle, finite-accuracy measurements, POVM}




\section{Introduction} \label{Sec.1}

In canonical quantum mechanics the Heisenberg uncertainty principle (HUP) implies a fundamental trade-off between the precisions of position and momentum measurements.
\footnote{In classical error analysis the term `precision' is used to refer to the statistical spread of the results whereas the term `accuracy' refers to the discrepancy between the measured value of a quantity and its true value. In keeping with this general usage, we use the term precision to refer to the quantum mechanical uncertainty and accuracy to refer to the width of the error bars associated with each individual measurement.} 
It can be introduced heuristically, via the famous Heisenberg microscope thought experiment, giving \cite{Heisenberg:1927zz,Heisenberg:1930}
\begin{eqnarray} \label{HUP-1}
\Delta x^{i} \, \Delta p_j \gtrsim \frac{\hbar}{2} \delta^{i}{}_{j} \, ,
\end{eqnarray}  
or derived rigorously from the canonical quantum formalism, yielding \cite{Rae,Isham}
\begin{eqnarray} \label{HUP-2}
\Delta_\psi x^i \, \Delta_\psi p_j  \geq \frac{\hbar}{2} \delta^{i}{}_{j} \, .
\end{eqnarray}
The inequality in Eq. (\ref{HUP-2}) is exact and, unlike the heuristic uncertainties $\Delta x^i$ and $\Delta p_j$ in Eq. (\ref{HUP-1}), $\Delta_\psi x^i$ and $\Delta_\psi p_j$ represent well-defined standard deviations of the probability distributions $|\psi(\bf{x})|^2$ and $|\tilde{\psi}_{\hbar}(\bf{p})|^2$, respectively, where the momentum space representation of the particle wave function is given by the $\hbar$-scaled Fourier transform of its position space representation: 
\begin{eqnarray} \label{dB-1}
\tilde{\psi}_{\hbar}({\bf p}) = \left(\frac{1}{\sqrt{2\pi\hbar}}\right)^3 \int \psi({\bf x}) e^{-\frac{i}{\hbar}{\bf p}.{\bf x}}{\rm d}^{3}{\rm x} \, .
\end{eqnarray}
We emphasise the scale-dependence of the canonical quantum Fourier transform, which is often neglected in standard treatments, by introducing the subscript $\hbar$. 
The relation (\ref{HUP-2}) is obtained by combining the Schr{\" o}dinger-Robertson relation for arbitrary Hermitian operators, $\hat{O}_1$ and $\hat{O}_2$ \cite{Robertson:1929zz,Schrodinger:1930ty}, 
\begin{eqnarray} \label{Robertson-Schrodinger-1}
\Delta_\psi O_1 \, \Delta_\psi O_2 \geq \frac{1}{2} |\langle \psi | [\hat{O}_1,\hat{O}_2] | \psi \rangle|  \, ,
\end{eqnarray} 
with the canonical position-momentum commutator,
\begin{eqnarray} \label{[x,p]-1}
[\hat{x}^i,\hat{p}_j] = i\hbar\delta^{i}{}_{j} \ \hat{\mathbb{I}} \, .
\end{eqnarray}

In recent years, thought experiments in quantum gravity research have suggested the existence of generalised uncertainty relations (GURs). 
By reconsidering Heisenberg's 1927 gedanken experiment, and accounting for the gravitational interaction between the massive particle and the probing photon, we obtain the generalised uncertainty principle (GUP), 
\begin{eqnarray} \label{GUP-1}
\Delta x^i \gtrsim \frac{\hbar}{2\Delta p_j} \delta^{i}{}_{j} \left[1 + \alpha_0 \frac{2G}{\hbar c^3}(\Delta p_j)^2\right] \, ,
\end{eqnarray} 
where $\alpha_0$ is a numerical constant of order unity \cite{Maggiore:1993rv,Adler:1999bu,Scardigli:1999jh}.
By minimising the right-hand side with respect to $\Delta p_j$,
the GUP implies the existence of a minimum position uncertainty of the order of the Planck length, $l_{\rm Pl} = \sqrt{\hbar G/c^3} \simeq 10^{-33} \, {\rm cm}$.

Reconsidering Heisenberg's arguments in the presence of a constant dark energy density $\rho_{\Lambda} = \Lambda c^2/(8\pi G) \simeq 10^{-30} \, {\rm g \, . \, cm^{-3}}$ \cite{Reiss1998,Perlmutter1999},
or, equivalently, an asymptotically de Sitter background with minimum scalar curvature of the order of the cosmological constant, $\Lambda \simeq 10^{-56} \, {\rm cm^{-2}}$ \cite{Betoule:2014frx,Ade:2013zuv}, gives the extended uncertainty principle (EUP), 
\begin{eqnarray} \label{EUP-1}
\Delta p_j \gtrsim \frac{\hbar}{2\Delta x^i} \delta^{i}{}_{j} \left[1 + 2\eta_0 \Lambda (\Delta x^i)^2\right] \, , 
\end{eqnarray} 
where $\eta_0$ is of order one \cite{Bolen:2004sq,Park:2007az,Bambi:2007ty}. 
The EUP implies the existence of a minimum momentum uncertainty of the order of the de Sitter momentum, $m_{\rm dS}c = \hbar\sqrt{\Lambda/3} \simeq 10^{-56} \, {\rm g \, . \, cm \, s^{-1}}$. 
This is physically reasonable since it is the minimum momentum that a canonical quantum particle can possess, when its wave function is localised within the asymptotic de Sitter horizon, which is comparable to the present day radius of the Universe $r_{\rm U}(t_0) \simeq l_{\rm dS} = \sqrt{3/\Lambda} \simeq 10^{28} \, {\rm cm}$. 

Combining both effects yields the extended generalised uncertainty principle (EGUP),
\begin{eqnarray} \label{EGUP-1}
\Delta x^i\Delta p_j \gtrsim \frac{\hbar}{2} \delta^{i}{}_{j} \left[1 + \alpha_0 \frac{2G}{\hbar c^3}(\Delta p_j)^2 + 2\eta_0\Lambda (\Delta x^i)^2\right] \, ,
\end{eqnarray} 
which implies the existence of both minimum length and momentum scales in nature \cite{Bolen:2004sq,Park:2007az,Bambi:2007ty}. 
Like their forebearer (\ref{HUP-1}) all three relations (\ref{GUP-1}), (\ref{EUP-1}) and (\ref{EGUP-1}) are heuristic in nature and it remains an open problem how to rigorously derive GURs from within a modified quantum formalism.  

Perhaps the simplest way to obtain the GUP, EUP or EGUP, given the relation (\ref{Robertson-Schrodinger-1}), is to modify the canonical position-momentum commutation relation (\ref{[x,p]-1}) and it is clear that a modification of the form 
\begin{eqnarray} \label{EGUP_[X,P]-1}
[\hat{x}^i,\hat{p}_j] = i\hbar\delta^{i}{}_{j} \ \hat{\mathbb{I}} \mapsto [\hat{X}^i,\hat{P}_j] = i\hbar\delta^{i}{}_{j} \left(\hat{\mathbb{I}} + \alpha_0 \frac{2G}{\hbar c^3} (\hat{P}_{j})^2 +  2\eta_0 \Lambda (\hat{X}^{i})^2 \right) 
\end{eqnarray}
gives rise to an EGUP-type uncertainty relation, at least when both $\langle \hat{P}_{j} \rangle_{\psi} = 0$ and $\langle \hat{X}^{i} \rangle_{\psi} = 0$ \cite{Kempf:1994su}. 
Here, we use capital letters to denote modified operators, which generate modified commutators, and lower case letters to denote their canonical quantum counterparts.
However, the assumption above is problematic since, even if both $\langle \hat{P}_{j} \rangle_{\psi} = 0$ and $\langle \hat{X}^{i} \rangle_{\psi} = 0$ in a given frame of reference, a simple shift of coordinate origin or a Galilean velocity boost of the observer alters the numerical value of the associated Schr{\" o}dinger-Robertson bound:
\begin{eqnarray} \label{EGUP-2}
\Delta_{\psi} X^i\Delta_{\psi} P_j \geq \frac{\hbar}{2} \delta^{i}{}_{j} \left\{1 + \alpha_0 \frac{2G}{\hbar c^3}[(\Delta_{\psi} P_j)^2 + \langle \hat{P}_j \rangle_{\psi}^2] + 2\eta_0\Lambda [(\Delta_{\psi} X^i)^2 + \langle \hat{X}^i \rangle_{\psi}^2]\right\} \, .
\end{eqnarray} 
This leads immediately to the reference frame-dependence of the (supposedly invariant) minimum length. 
In fact, the situation is even worse since even a redefinition of the position-coordinate origin alters the value of the bound on the right-hand side. 
This gives rise to a coordinate-dependent `minimum' length, which is clearly unphysical, and which strongly suggests that GUR models based on modified commutation relations are not mathematically self-consistent \cite{Lake-Book-Chapter,Lake-Special-Issue}

In addition, the modified position-momentum commutator (\ref{EGUP_[X,P]-1}) implies a modification of the canonical Heisenberg equation, which immediately gives rise to mass-dependent accelerations for quantum particles, violating the equivalence principle \cite{Tawfik:2014zca,Tawfik:2015rva}.
Such models also violate Lorentz invariance in the relativistic limit and suffer from the so-called soccer ball problem, so that sensible GUP-compatible multi-particle states cannot be defined \cite{Hossenfelder:2012jw,Amelino-Camelia:2014gga}.
\footnote{In \cite{Amelino-Camelia:2014gga} 
an ingenious solution to the soccer ball problem was proposed by Amelino-Camelia.   
In this approach, the generalised momentum operators of a given modified commutator model are defined to be the generators of `generalised spatial translations'. 
The unitary transformation $\hat{\mathcal{U}}({\bf X}) := \exp[(i/\hbar){\bf X}.\hat{{\bf P}}]$, which acts nontrivially only on the $\hat{X}^{i}$ operators, is required to leave the modified $[\hat{X}^{i},\hat{P}_{j}]$, $[\hat{X}^{i},\hat{X}^{j}]$ and $[\hat{P}_{i},\hat{P}_{j}]$ algebras, as well as the multi-particle Hamiltonian of the model, $\hat{\mathcal{H}}$, invariant. This defines the `generalised translation symmetries' of the system and, when these symmetries hold, the corresponding Noether charge for an $N$-particle state is represented by the operator $\hat{{\bf P}}_{\rm Total} := \sum_{I=1}^{N}\hat{{\bf P}_{I}}$, where $[\hat{{\bf P}}_{\rm Total} ,\hat{\mathcal{H}}] = 0$. The usual law of linear momentum addition therefore holds for multi-particle states but a different nonlinear addition law, derived ultimately from the notion of spatial locality, holds for transfers of momentum between individual particles, due to the interactions specified by $\hat{\mathcal{H}}$ \cite{Amelino-Camelia:2014gga}.
Unfortunately for GUP models, in the example system considered in \cite{Amelino-Camelia:2014gga}, 
the definition of the generalised spatial translations required to maintain the linear addition law also requires one of the position-momentum commutators to equal zero, i.e., $[\hat{X}^i,\hat{P}_i] = 0$, for some $i$. 
In this case there is no Heisenberg uncertainty principle, let alone a GUP, even though a minimum length scale $l$ still appears in the model via the position-position commutator, e.g. $[\hat{X}_1,\hat{X}_2] = il\hat{X}_1$. 
This illustrates a general point, that it is by no means certain whether a particular modified momentum operator, corresponding to a particular modification of the canonical Heisenberg algebra, and, hence, a particular form of the GUP, is compatible with a linear addition law derived via Amelino-Camelia's procedure. 
Therefore, although this procedure represents a useful criterion for defining physically viable GUP models, it is clear that arbitrary deformations of the canonical Heisenberg algebra are not consistent with the existence of a linear momentum addition law and that further work is required to determine which models truly suffer from a soccer ball problem and which ones do not. 
Though some GUP models may be free from this pathology, a great many could still be afflicted by it.}

The heuristic, model-independent nature of the gedanken experiments that lead to the relations (\ref{GUP-1}), (\ref{EUP-1}) and (\ref{EGUP-1}), together with the pathologies displayed by modified commutator models, motivate us to consider alternative ways to generate GUP, EUP and EGUP phenomenology, without modifying the canonical Heisenberg algebra. 
In this paper, we consider one way in which such a scheme can be implemented from {\it within} the canonical quantum formalism. 
The physical basis of the model is the notion of a finite-accuracy measurement and these are represented mathematically by the construction of appropriate POVM. 
Roughly speaking, since errors add in quadrature for independent random variables, finite-accuracy measurements of position and momentum with detection `sweet spots' of width $\sigma_g \simeq l_{\rm Pl}$ and $\tilde{\sigma}_g \simeq m_{\rm dS}c$, respectively, give rise to the GUP and EUP, to first order in the relevant Taylor expansion. 
These individual relations may then be combined to give the EGUP. 


\section{GUR from finite-accuracy measurements described by POVM} \label{Sec.2}

In this section, we show that GUP, EUP and EGUP-type uncertainty relations can be derived in an effective model, where position and momentum measurements in {\it canonical quantum theory} are not perfectly accurate and are described by POVMs, rather than perfect projective measurements. 

Let us begin by replacing the usual position-measurement operator, $\hat{{\bf x}}$, with POVM elements corresponding to the result ${\bf x}$:
\begin{equation} \label{E_x}
\hat{E}_{\bf{x}} := \int g({\bf x}' - {\bf x}) |{\bf x}'\rangle \langle{{\bf x}'}|{\rm d}^{3}{\rm x}' \, ,
\end{equation}
where $g({\bf x}' - {\bf x})$ is any normalised function, $\int |g({\bf x}' - {\bf x})|^2{\rm d}^{3}{\rm x}' = 1$.
These elements satisfy the relations $\hat{E}_{{\bf x}}^\dagger \hat{E}_{{\bf x}} \ge 0$ and $\int \hat{E}_{{\bf x}}^\dagger \hat{E}_{{\bf x}} {\rm d}^3{\rm x} = \hat{\mathbb{I}}$, as required, so that Eq. (\ref{E_x}) defines a standard POVM in canonical quantum mechanics \cite{Chuang_Nielsen}.
From here on, we refer to $g$ as the `envelope function' of the measure. 
For spherically symmetric functions the envelope is centred on the value $\bf{x}$, and, for the sake of concreteness, we may imagine $|g({\bf x}' - {\bf x})|^2$ as a three-dimensional Gaussian distribution with mean ${\bf x}$ and standard deviation $\sigma_g$.

Finite-accuracy position measurements, conducted on an arbitrary state $|\psi\rangle$, then give rise to the first and second order moments
\begin{eqnarray}
\langle E_{{\bf x}} \rangle_\psi &=& \int {\bf x} \langle \psi | \hat{E}_{{\bf x}}^\dagger \hat{E}_{{\bf x}} | \psi \rangle {\rm d}^3{\rm x} =  \langle {\bf x} \rangle_{g} + \langle {\bf x} \rangle_\psi \, , 
\nonumber\\
\langle E_{{\bf x}}^2 \rangle_\psi &=& \int {\bf x}^2 \langle \psi | \hat{E}_{{\bf x}}^\dagger \hat{E}_{{\bf x}} | \psi \rangle {\rm d}^3{\rm x}  =  \langle {\bf x}^2 \rangle_{g} + \langle {\bf x}^2 \rangle_\psi \, , 
\end{eqnarray}
where $\langle {\bf x}^n \rangle_{f} := \int {\bf x}^n \, |f({\bf x})|^2 \, {\rm d}^3{\rm x}$ with $f({\bf x}) = g({\bf x})$ or $\psi({\bf x})$. 
Since $|g({\bf x}'-{\bf x})|^2$ is a normalised function centred on ${\bf x}'={\bf x}$, $\langle {\bf x} \rangle_{g} = 0$, and the corresponding variance is given by
\begin{eqnarray} \label{x-var-1}
(\Delta_\psi E_{{\bf x}})^2 = (\Delta_\psi {\bf x})^2 + \boldsymbol{\sigma}_g^2 \, ,
\end{eqnarray}
where $\boldsymbol{\sigma}_g :=$$\, \sigma_g^{i}$${\bf e}_{i}$ and $\sigma_g^{i}$ denotes the width of $|g|^2$ in each coordinate direction $x^{i}$. 
By spherical symmetry, $\sigma_g^{i} =  \sigma_g$ for all $i$, and we may rewrite Eq. (\ref{x-var-1}) in terms of the individual components as
\begin{eqnarray} \label{x-var-2}
(\Delta_\psi E_{i})^2 = (\Delta_\psi x^{i})^2 + \sigma_g^2 \, ,
\end{eqnarray}
where we have used the shorthand notation $\Delta_\psi E_{i} \equiv \Delta_\psi E_{x^{i}}$.

In like manner, finite-accuracy momentum measurements may be introduced via the operators
\begin{equation} \label{E_p}
\hat{\mathbb{E}}_{{\bf p}} := \int \tilde{g}({\bf p}' - {\bf p}) |{\bf p}'\rangle\langle{\bf p'}| {\rm d}^{3}{\rm p}' \, ,
\end{equation}
where $\int |\tilde{g}({\bf p}' - {\bf p})|^2 {\rm d}{\rm p}'= 1$, but it is important to note that there is no {\it intrinsic} relation between the functions $g$ and $\tilde{g}$, which may be chosen independently for a given POVM model. 
Nevertheless, if both $|g|^2$ and $|\tilde{g}|^2$ represent Gaussian distributions, which is perhaps the most natural choice for an envelope function, then $g$ and $\tilde{g}$ {\it are} related via a Fourier transform,
\begin{equation} \label{g_Fourier}
\tilde{g}({\bf p}' - {\bf p}) = \int g({\bf x}' - {\bf x}) e^{\frac{i}{\beta}({\bf x}' - {\bf x}).({\bf p}' - {\bf p})} {\rm d}^{3}{\rm x}' \, , 
\end{equation}
where the new action scale $\beta \neq \hbar$ is given by
\begin{equation} \label{beta}
\beta := 2\sigma_g\tilde{\sigma}_g \, , 
\end{equation}
and $\tilde{\sigma}_g$ is the standard deviation of $|\tilde{g}|^2$. 
However, it is equally important to note that there is nothing fundamental about the relation (\ref{g_Fourier}). 
Unlike the $\hbar$-scaled Fourier transform relating the position and momentum space representations of the quantum wave function, Eq. (\ref{dB-1}), the $\beta$-scaled transform relates the `envelope functions' of the model.

Finite-accuracy momentum measurements, conducted on an arbitrary state $|\psi\rangle$, then give rise to the first and second order moments
\begin{eqnarray}
\langle \mathbb{E}_{{\bf p}} \rangle_\psi & = &  \int {\bf p} \langle \psi | \hat{\mathbb{E}}_{{\bf p}}^\dagger \hat{\mathbb{E}}_{{\bf p}} | \psi \rangle {\rm d}^3{\rm p} = \langle {\bf p} \rangle_{g} + \langle {\bf p} \rangle_\psi \, , 
\nonumber\\
\langle \mathbb{E}_{{\bf p}}^2 \rangle_\psi & = &  \int {\bf p}^2 \langle \psi | \hat{\mathbb{E}}_{{\bf p}}^\dagger \hat{\mathbb{E}}_{{\bf p}} | \psi \rangle {\rm d}^3{\rm p}  = \langle {\bf p}^2 \rangle_{g} + \langle {\bf p}^2 \rangle_\psi \, , 
\end{eqnarray}
where $\langle {\bf p}^n \rangle_{f} := \int {\bf p}^n \, |\tilde{f}({\bf p})|^2 \, {\rm d}^3{\rm p}$ with $\tilde f({\bf p}) = \tilde g(\bf{p})$ or $\tilde \psi_\hbar (\bf{p})$. 
Since $|\tilde{g}({\bf p}'-{\bf p})|^2$ is normalised and centred at ${\bf p}' = {\bf p}$, $\langle {\bf p} \rangle_{g} = 0$, and 
\begin{eqnarray} \label{p-var-1}
(\Delta_\psi \mathbb{E}_{{\bf p}})^2 = (\Delta_\psi {\bf p})^2 + \boldsymbol{\tilde{\sigma}}_g^2 \, ,
\end{eqnarray}
where $\boldsymbol{\tilde{\sigma}}_g := $$\, \tilde{\sigma}_{gj}$${\bf e}^{j}$ and $\tilde{\sigma}_{gj}$ denotes the width of $|\tilde{g}|^2$ in each momentum space direction $p_{j}$. 
Again employing spherical symmetry, $\tilde{\sigma}_{gj} = \tilde{\sigma}_{g}$ for all $j$, so that (\ref{p-var-1}) may be rewritten in terms of the individual components as
\begin{eqnarray} \label{p-var-2}
(\Delta_\psi \mathbb{E}_{j})^2 = (\Delta_\psi p_{j})^2 + \tilde{\sigma}_g^2 \, ,
\end{eqnarray}
where we have again used the shorthand $\Delta_\psi \mathbb{E}_{j} \equiv \Delta_\psi \mathbb{E}_{p_j}$. 

To obtain a GUP-type relation from Eq. (\ref{x-var-2}) we simply take the square root, Taylor expand the right-hand side to first order, and substitute for $\Delta_{\psi} x^{i}$ from the HUP (\ref{HUP-2}). 
Likewise, an EUP-type relation is obtained from (\ref{p-var-2}) by taking the square root, Taylor expanding to first order, and substituting for $\Delta_{\psi} p_{j}$. 
Next, using the substitutions
\begin{eqnarray} \label{}
\sigma_g := \sqrt{2\alpha_0} \, l_{\rm Pl} \, , \quad \tilde{\sigma}_g := \sqrt{6\eta_0} \, m_{\rm dS}c \, , 
\end{eqnarray}
where
\begin{eqnarray} \label{}
l_{\rm Pl} :=  \sqrt{\hbar G/c^3} \, , \quad m_{\rm dS}c := \hbar \sqrt{\Lambda/3} \, , 
\end{eqnarray}
immediately gives
\begin{eqnarray} \label{GUP-2}
\Delta_{\psi} X^i \gtrsim \frac{\hbar}{2\Delta_{\psi} p_j} \delta^{i}{}_{j} \left[1 + \alpha_0 \frac{2G}{\hbar c^3}(\Delta_{\psi} p_j)^2\right] \, ,
\end{eqnarray} 
\begin{eqnarray} \label{EUP-2}
\Delta_{\psi} P_j \gtrsim \frac{\hbar}{2\Delta_{\psi} x^i} \delta^{i}{}_{j} \left[1 + 2\eta_0 \Lambda (\Delta_{\psi} x^i)^2\right] \, , 
\end{eqnarray} 
where we have relabelled $\Delta_{\psi}E_{i} \equiv \Delta_{\psi}X^i$ and $\Delta_{\psi}\mathbb{E}_{j} \equiv \Delta_{\psi}P_j$, for convenience. 
These expressions are formally analogous to the heuristic relations (\ref{GUP-1}) and (\ref{EUP-1}), respectively, but with $\Delta p_j$ and $\Delta x^i$ on the right replaced by the well-defined standard deviations $\Delta_{\psi} p_j$ and  $\Delta_{\psi} x^i$. 

This proves that GUP- and EUP-type relations can be derived rigorously, from within the canonical quantum formalism, but a remaining criticism of the formulae above is that the uncertainties on the right-hand sides of Eqs. (\ref{GUP-2})-(\ref{EUP-2}) are not equivalent to the uncertainties on the left. 
Indeed, according to the POVM model, $\Delta_{\psi} p_j$ and $\Delta_{\psi} x^i$ are not operationally {\it observable} quantities. 
They arise only in the limits $\sigma_g \rightarrow 0$ and $\tilde{\sigma}_g \rightarrow 0$, respectively, in which both (\ref{GUP-2}) and (\ref{EUP-2}) reduce to the standard HUP (\ref{HUP-2}). 
This objection can be overcome, however, by first substituting for $\Delta_{\psi} x^i$ from (\ref{HUP-2}) in Eq. (\ref{x-var-2}) and then again for $\Delta_{\psi} p_i$ from (\ref{p-var-2}). 
This gives rise to an uncertainty relation between the observable standard deviations, $\Delta_{\psi}E_{i} \equiv \Delta_{\psi}X^i$ and $\Delta_{\psi}\mathbb{E}_{j} \equiv \Delta_{\psi}P_j$. 
It is straightforward to show that, taking the square root, Taylor expanding to first order, and neglecting the final term of order $\sigma_g\tilde{\sigma}_g \simeq l_{\rm Pl} \, . \, m_{\rm dS}c$, this relation reduces to
\begin{eqnarray} \label{EGUP-2}
\Delta_{\psi} X^i\Delta_{\psi} P_j \gtrsim \frac{\hbar}{2} \delta^{i}{}_{j} \left[1 + \alpha_0 \frac{2G}{\hbar c^3}(\Delta_{\psi} P_j)^2 + 2\eta_0\Lambda (\Delta_{\psi} X^i)^2\right] \, .
\end{eqnarray} 
Therefore, the EGUP can be rigorously derived within the canonical quantum formalism. 
The GUP and EUP proper then arise as limits of this more fundamental relation.  

We stress that, in this model, $\Delta_{\psi}E_{i} \equiv \Delta_{\psi}X^i$ and $\Delta_{\psi}\mathbb{E}_{j} \equiv \Delta_{\psi}P_j$ represent the {\it physically observable} precisions, obtained from generalised position and momentum measurements with finite accuracies $\sigma_g > 0$ and $\tilde{\sigma}_g > 0$. 
By contrast, the canonical Hamiltonian is determined by the canonical (projective) position and momentum operators, $\hat{\bf{x}}$ and $\hat{\bf{p}}$, via $\hat{H} = \hat{\bf{p}}^2/(2m) + V(\hat{\bf{x}})$, where the former obey the canonical Heisenberg algebra: $[\hat{x}^i,\hat{p}_j] = i\hbar\delta^{i}{}_{j} \ \hat{\mathbb{I}}$, $[\hat{x}^i,\hat{x}^j] = 0$,  $[\hat{p}_i,\hat{p}_j] = 0$. 
This leaves the canonical Heisenberg and Schr{\" o}dinger equations unchanged and neatly evades the pathologies that afflict modified commutator models \cite{Lake-Book-Chapter,Lake-Special-Issue,Tawfik:2014zca,Tawfik:2015rva,Hossenfelder:2012jw}.

\section{Discussion} \label{Sec.3}

We have shown that the three most common GURs studied in the quantum gravity literature, the GUP, EUP and EGUP, can be derived from within the formalism of canonical quantum mechanics. 
A GUP-type uncertainty relation is obtained when the standard (projective) position operator is replaced by an appropriate POVM, representing finite-accuracy measurements with error bars of width $\sigma_g > 0$ in real space. 
In like manner, an EUP-type relation is obtained from finite-accuracy measurements with error bars of width $\tilde{\sigma}_g > 0$ in momentum space. 
These can be combined to give a relation that is formally analogous to the EGUP and the standard EGUP is recovered by setting $\sigma_g \simeq l_{\rm Pl}$, the Planck length, and $\tilde{\sigma}_g \simeq m_{\rm dS}c$, where $m_{\rm dS} = (\hbar/c)\sqrt{\Lambda/3}$ is the de Sitter mass. 

This work suggests that GUP, EUP and EGUP phenomenology can be understood in a physically intuitive way, as a simple and natural outcome of finite-accuracy measurements. 
Such measurements are capable of generating all three GURs and the same phenomenology is obtained, at the level of the uncertainty relations, regardless of whether the limits $(\Delta_{\psi}X^{i})_{\rm min} = \sigma_g$ and 
$(\Delta_{\psi}P_{j})_{\rm min} = \tilde{\sigma}_g$ are fundamental, or merely effective, as an outcome of an imperfect measurement scheme. 

We propose that this should give pause for thought to the GUP community. 
If modified commutators are not {\it necessary} for GUP phenomenology, and, after nearly 30 years of research, we are no closer to resolving the pathologies that have afflicted these models since they were first proposed in the mid-1990's, then serious attempts should be made to find {\it alternative mathematical structures} that give rise to GURs. 
These should be capable of generating, via rigorous derivation, the uncertainty relations predicted by model-independent gedanken experiments, but without the problems associated with modified commutation relations. 

In this paper, we have proposed one such model, within the context of canonical quantum theory. 
Another, more radical, alternative is to consider additional quantum mechanical degrees of freedom, not present in the canonical theory, which are capable of describing quantum fluctuations of the background geometry. 
Such a model was proposed in a recent series of works \cite{Lake:2018zeg,Lake:2019oaz,Lake:2019nmn,Lake:2021beh,Lake:2020chb}
and shares many features with the model described here, including the existence of a new action scale that relates the accuracies of generalised position and momentum measurements, $\beta := 2\sigma_g\tilde{\sigma}_g \simeq 10^{-61}\hbar$ (*). 
The fundamental difference between the two models is the existence of new degrees of freedom in the latter. 
From this, it follows that the new action scale $\beta$ implies a modified de Broglie relation of the form $\bf{p}' = \hbar\bf{k} + \beta(\bf{k}' - \bf{k})$, where, here, $\bf{p}'$ denotes the {\it observable} momentum. 
Heuristically, the non-canonical term $\beta(\bf{k}' - \bf{k})$ can be interpreted as an additional momentum `kick', transferred to the canonical wave function by a quantum fluctuation of the background. 
The interested reader is referred to \cite{Lake-Book-Chapter,Lake-Special-Issue,Lake:2018zeg,Lake:2019oaz,Lake:2019nmn,Lake:2021beh,Lake:2020chb}
for further details. 

At first glance, this more radical alternative has nothing to do with the POVM approach described here. It requires extra degrees of freedom associated with the quantum state of the background geometry, contrary to the POVM formalism, which remains entirely within the context of canonical quantum theory. It follows from Stinespring's dilation theorem \cite{Stinespring:1955, Paulson_Book}, however, that the two formalisms are equivalent if we assume the particular values, $\sigma_g \simeq l_{\rm Pl}$ and $\tilde{\sigma}_g \simeq m_{\rm dS}c$, and hence the relation (*) above. 
The POVM picture results from tracing out the ${\bf x}'$ (${\bf p}'$) degrees of freedom associated with quantum fluctuations of the background and the ${\bf x}'$ (${\bf p}'$) degrees of freedom appear as a consequence of dilating the POVM.

The POVM approach describes a quantum measurement of finite accuracy. 
The minimum resolution of the measurement may be due to technical limitations, or it can reflect the fact that the minimum length and
momentum scales are fundamentally related. We postulate that in a universe with both fundamental and technological limitations to measurement accuracy, the complete description of a realistic quantum measurement should be a POVM extension of the model presented in \cite{Lake:2018zeg, Lake:2019oaz}. 
We expect that this would give rise to two additional contributions to the position and momentum variances, i.e., $ \sigma_g^2 + \sigma_h^2$ and $\tilde{\sigma}_g^2 + \tilde{\sigma}_h^2$, respectively, where $g$ is the fundamental smearing function that models the quantum indeterminacy of space-time, and $h$ is the envelope function of a realistic detector. In the limit $\sigma_h \gg \sigma_g$, $ \tilde{\sigma}_h \gg \tilde{\sigma}_g$, which corresponds to all present-day measurements, the latter are expected to dominate the former.

\section*{Acknowledgments}

ML acknowledges the Department of Physics and Materials Science, Faculty of Science, Chiang Mai University, for providing research facilities, and the Natural Science Foundation of Guangdong Province, grant no. 008120251030.





\begin{thebibliography}{99}


\bibitem{Heisenberg:1927zz} 
  W.~Heisenberg,
  {\it {\" U}ber den anschaulichen Inhalt der quantentheoretischen Kinematik und Mechanik},
  Z.\ Phys.\  {\bf 43}, 172 (1927).

\bibitem{Heisenberg:1930}  
   W.~Heisenberg, 
   {\it The physical principles of the quantum theory}, 
   New York, Dover (1930).
 
\bibitem{Rae}
   A.~I.~M.~Rae,  
   {\it Quantum Mechanics}, 4th  ed.,
   Taylor \& Francis:  London, U.K. (2002). 
    
\bibitem{Isham}
   C.~J.~Isham, 
   {\it Lectures on Quantum Theory: Mathematical and Structural Foundations},
   Imperial College Press, London (1995).   
 
\bibitem{Robertson:1929zz} 
   H.~P.~Robertson,
   {\it The Uncertainty Principle},
   Phys.\ Rev.\  {\bf 34}, 163 (1929).
 
\bibitem{Schrodinger:1930ty} 
   E.~Schr{\" o}dinger,
   {\it About Heisenberg uncertainty relation},
   Bulg.\ J.\ Phys.\  {\bf 26}, 193 (1999)
   [Sitzungsber.\ Preuss.\ Akad.\ Wiss.\ Berlin (Math.\ Phys.\ ) {\bf 19}, 296 (1930)].


\bibitem{Maggiore:1993rv} 
   M.~Maggiore,
   {\it A Generalized uncertainty principle in quantum gravity},
   Phys.\ Lett.\ B {\bf 304}, 65 (1993).
  
\bibitem{Adler:1999bu} 
   R.~J.~Adler and D.~I.~Santiago,
   {\it On gravity and the uncertainty principle},
   Mod.\ Phys.\ Lett.\ A {\bf 14}, 1371 (1999).

\bibitem{Scardigli:1999jh} 
   F.~Scardigli,
   {\it Generalized uncertainty principle in quantum gravity from micro - black hole Gedanken experiment},
   Phys.\ Lett.\ B {\bf 452}, 39 (1999).
   
  
\bibitem{Reiss1998} 
   A. G. Riess {\it et al.}, 
   {\it Observational Evidence from Supernovae for an Accelerating Universe and a Cosmological Constant}, 
   Astron. J. \textbf{116}, 1009 (1998).

\bibitem{Perlmutter1999} 
   S. Perlmutter {\it et al.}, 
   {\it Measurements of $\Omega$ and $\Lambda$ from 42 high-redshift supernovae}, 
   Astrophys. J. \textbf{517}, 565 (1999).

  
\bibitem{Betoule:2014frx} 
   M.~Betoule {\it et al.} [SDSS Collaboration],
   {\it Improved cosmological constraints from a joint analysis of the SDSS-II and SNLS supernova samples},
   Astron.\ Astrophys.\  {\bf 568}, A22 (2014).
  

\bibitem{Ade:2013zuv} 
   P.~A.~R.~Ade {\it et al.} [Planck Collaboration],
   {\it Planck 2013 results. XVI. Cosmological parameters},
   Astron.\ Astrophys.\  {\bf 571}, A16 (2014).

  
\bibitem{Bolen:2004sq} 
   B.~Bolen and M.~Cavaglia,
   {\it (Anti-)de Sitter black hole thermodynamics and the generalized uncertainty principle},
   Gen.\ Rel.\ Grav.\  {\bf 37}, 1255 (2005).
    
\bibitem{Park:2007az} 
   M.~i.~Park,
   {\it The Generalized Uncertainty Principle in (A)dS Space and the Modification of Hawking Temperature from the Minimal Length},
   Phys.\ Lett.\ B {\bf 659}, 698 (2008).
  
\bibitem{Bambi:2007ty} 
   C.~Bambi and F.~R.~Urban,
   {\it Natural extension of the Generalised Uncertainty Principle},
   Class.\ Quant.\ Grav.\  {\bf 25}, 095006 (2008).

 
\bibitem{Kempf:1994su} 
   A.~Kempf, G.~Mangano and R.~B.~Mann, 
  {\it Hilbert space representation of the minimal length uncertainty relation},
   Phys. Rev. D \textbf{52}, 1108-1118 (1995).

 
\bibitem{Lake-Book-Chapter}
   M.~J.~Lake,
   {\it A New Approach to Generalised Uncertainty Relations}, 
   to appear in Touring the Planck scale: Antonio Aurilia memorial volume, Piero Nicolini, ed., 
   Fundamental Theories of Physics, Springer 
   [arXiv:2008.13183 [gr-qc]] (2020).   
   

\bibitem{Lake-Special-Issue}
   M.~J.~Lake and A.~Watcharapasorn,
   {\it Problems with modified commutators},  
   Front. Astron. Space Sci., {\bf 10}, 1118647 (2023).
   
  
\bibitem{Tawfik:2014zca} 
   A.~N.~Tawfik and A.~M.~Diab,
   {\it Generalized Uncertainty Principle: Approaches and Applications},
   Int.\ J.\ Mod.\ Phys.\ D {\bf 23}, no. 12, 1430025 (2014).
   doi:10.1142/S0218271814300250
   [arXiv:1410.0206 [gr-qc]].
  
\bibitem{Tawfik:2015rva} 
    A.~N.~Tawfik and A.~M.~Diab,
   {\it Review on Generalized Uncertainty Principle},
   Rept.\ Prog.\ Phys.\  {\bf 78}, 126001 (2015)
   
\bibitem{Hossenfelder:2012jw} 
   S.~Hossenfelder,
   {\it Minimal Length Scale Scenarios for Quantum Gravity},
   Living Rev.\ Rel.\  {\bf 16}, 2 (2013).
       
  
\bibitem{Amelino-Camelia:2014gga}
   G.~Amelino-Camelia,
   {\it Planck-scale soccer-ball problem: a case of mistaken identity},
   Entropy \textbf{19}, no.8, 400 (2017).


\bibitem{Chuang_Nielsen}
   M.~A.~Nielsen and I.~L.~Chuang, 
   {\it Quantum Computation and Quantum Information}, 
   Cambridge University Press, Cambridge (2000). 


\bibitem{Lake:2018zeg}
   M.~J.~Lake, M.~Miller, R.~F.~Ganardi, Z.~Liu, S.~D.~Liang and T.~Paterek,
   {\it Generalised uncertainty relations from superpositions of geometries},
   Class.\ Quant.\ Grav.\  {\bf 36}, no.15,  155012 (2019).

\bibitem{Lake:2019oaz} 
   M.~J.~Lake,
   {\it A Solution to the Soccer Ball Problem for Generalized Uncertainty Relations},
   Ukr.\ J.\ Phys.\  {\bf 64}, no. 11, 1036 (2019).

\bibitem{Lake:2019nmn} 
   M.~J.~Lake, M.~Miller and S.~D.~Liang,
   {\it Generalised uncertainty relations for angular momentum and spin in quantum geometry},
   Universe 2020, 6, 56 (2020).
  
\bibitem{Lake:2021beh}
   M.~J.~Lake,
   {\it How Does the Planck Scale Affect Qubits?},
   Quantum Rep. \textbf{3}, no.1, 196-227 (2021).
  
\bibitem{Lake:2020chb}
    M.~J.~Lake,
    {\it Why space could be quantised on a different scale to matter,}
    SciPost Phys. Proc. \textbf{4}, 014 (2021).
    
\bibitem{Stinespring:1955}    
    W.~F.~Stinespring, 
    {\it Positive Functions on C*-algebras}, 
    Proceedings of the American Mathematical Society, 6, 211–216 (1955).
 
\bibitem{Paulson_Book}   
    V.~Paulsen, 
    {\it Completely Bounded Maps and Operator Algebras}, 
    Cambridge University Press, Cambridge, U.K. (2003).


\end{thebibliography}
\end{document}